\documentclass{epl}

\usepackage{amsfonts}
\usepackage{amssymb}
\usepackage{amsmath}

\newcommand{\dDM}{\delta_{\rm {DM}}}

\renewcommand{\vec}[1]{{\bf \boldsymbol #1}}

\newcommand{\QQ}{\vec{q}_0}

\title{Weak spin-orbit interactions induce exponentially flat mini-bands
  in magnetic metals without inversion symmetry}
\shorttitle{Spin-orbit interactions induce flat mini-bands}
\author{I. Fischer \and A. Rosch}
\institute{Institut f\"ur Theoretische Physik, 
Universit\"at zu K\"oln, 50937 K\"oln, Germany
}


\begin{document}

\maketitle

\begin{abstract}
  In metallic magnets like MnSi the interplay of two very weak
  spin-orbit coupling effects can strongly modify the Fermi surface. In the
  absence of inversion symmetry even a very small Dzyaloshinsky-Moriya
  interaction of strength $\delta \ll 1$ distorts a ferromagnetic
  state into a chiral helix with a long pitch of order $1/\delta$. We
  show that additional small spin-orbit coupling terms of order
  $\delta$ in the band structure lead to the formation of exponentially
  flat mini-bands with a bandwidth of order $ e^{-1/\sqrt{\delta}}$
  parallel to the direction of the helix. These flat mini-bands cover a
   rather broad belt of width $\sqrt{\delta}$ on the Fermi surface
  where electron motion parallel to the helix practically stops. We
  argue that these peculiar band-structure effects lead to pronounced
  features in the anomalous skin effect.
\end{abstract}

\section{Introduction} MnSi is a textbook 
example\cite{landau} of a magnetic metal where weak spin-orbit (SO)
coupling leads to a helical modulation of the ferromagnetic
order\cite{bak,earlyExp}. While this physics has been well understood
for almost 25 years\cite{bak}, the interest in MnSi was recently
renewed \cite{pfleiderer,lonzarich,nature} after it was discovered
\cite{pfleiderer} that moderate pressures $p$ suppress the long range
helical order and drive the system for $p>p_c\approx 14.6$kbar into a
novel state characterized by an anomalous resistivity, $\rho \sim
T^{3/2}$, which is observed over almost three decades\cite{pfleiderer}
in temperature $T$ and in a huge pressure range\cite{lonzarich}. This
has been taken as evidence for the existence of a genuine non-Fermi
liquid phase in this extremely clean cubic system. Recent neutron
scattering results\cite{nature} for $p>p_c$ suggest that while true
long-range order is lost in this phase, a peculiar partial helical
order survives on intermediate time and length scales.

This motivates us to study the motion of electrons in a helical
state in more detail. Within this paper we try to answer a simple
question: What is the generic band-structure of electrons in the
magnetically ordered state of a metal without inversion symmetry?  We
will show that even in the case of weak spin-orbit interaction (as
realized in MnSi) the answer to this question is surprisingly complex:
the interplay of  two weak spin-orbit effects of similar strength leads
to a pronounced restructuring of the Fermi surface. For a large
portion of the Fermi surface the electron motion parallel to the helix
turns out to be almost completely frozen. 

In the following we will briefly review how the Dzyaloshinsky-Moriya
interaction leads to helical order in magnets lacking inversion
symmetries. We then calculate the band structure within the
magnetically ordered phase first neglecting all spin-orbit effects
besides the Dzyaloshinsky-Moriya interaction. In this case, the
mini-bands, which formally arise in the reduced Brillouin zone of the helical
phase, can be avoided by a specific choice of reference frame.
In a second step we include leading spin-orbit corrections
in the band-structure which drastically modify this picture. Finally,
we discuss experimental consequences of our results.

\section{Formation of helix} In 1980 Nakanishi {\it et al.} and Bak and
H{\o}gh \cite{bak} investigated the Ginzburg-Landau theory of magnetic
systems like MnSi which lack an inversion symmetry.  The cubic metal
MnSi is characterized by the space group P2$_1$3 whose point group $T$
consists only of cyclic permutations of $\hat{x}$,$\hat{y}$ and
$\hat{z}$, of rotations by $\pi$ around the coordinate axes and
of combinations thereof. Spin orbit coupling is very weak in this system
(see below). Throughout this paper we will therefore consistently use
the strength of spin-orbit coupling (defined below) as a small
parameter. At low temperature MnSi has a sizable magnetic moment of
about $0.4\, \mu_B$. Therefore a Ginzburg-Landau expansion in the
amplitude of the order parameter is not possible, however, one can
still expand in the strength of spin-orbit coupling and gradients of
the order parameter $\vec{\Phi}(\vec{x})$
\begin{equation}\label{FM}
 \int \frac{1}{2}\sum_{ij} (\nabla_i \Phi_j)(\nabla_i \Phi_j)+q_0 \vec{\Phi}(\vec{x}) \cdot (\vec{\nabla} \times \vec{\Phi}(\vec{x})).
\end{equation}
The second term, the Dzyaloshinsky-Moriya interaction, arises in the
absence of inversion symmetries in {\em linear} order of spin orbit
coupling.
As it is {\em linear} in momentum, an arbitrary small $q_0$ destabilizes  the 
ferromagnetic state, twisting it into a helix of the form
 \begin{eqnarray}\label{phi0}
\vec{\Phi}(\vec{x})&=&\Phi_0 (\hat{\vec{n}}_1 \cos \QQ \vec{x}+\hat{\vec{n}}_2 \sin \QQ \vec{x})
\end{eqnarray}
where $\hat{\vec{n}}_1 \perp \hat{\vec{n}}_2 \perp \QQ$ are three
perpendicular vectors and $|\QQ|=q_0$. The weakness of relativistic
effects leads to a large pitch $2 \pi/q_0$ of the helix, 175\,\AA\ in
the case of MnSi.  The dimensionless constant $\dDM=q_0/k_F$, where
$k_F$ is a typical Fermi momentum, is therefore maximally of the order
of a few percent. The energy gain due to the formation of the helix is
of order $\dDM^2$, higher order corrections of order $\dDM^4$ lock  the
direction of the helix to the $\langle 111\rangle$ direction. While these terms are important for the
Goldstone modes in the system \cite{roschToBe}, they can be neglected
for the following discussion.

\section{Band structure without spin-orbit interaction} The large pitch of
the helix implies a large unit-cell in the ordered phase with more
than 300 atoms which makes a band-structure calculation from first
principles difficult. To our knowledge only non-relativistic
calculations\cite{band,deHaas} exist; those assume a
ferromagnetic state and neglect the helical modulation. The results are rather complex with several bands crossing the
Fermi energy consistent with  de Haas-van Alphen experiments \cite{deHaas} in large magnetic fields. We will not try to model these details but focus our
interest on qualitative features in the band-structure, for example we
will not keep track of multiple bands.  We therefore consider the
following simple non-interacting one-band Hamiltonian
\begin{equation}\label{h}
H= \sum_{\vec{k},\alpha \alpha'} 
\epsilon_{\vec{k}}^{\alpha \alpha'} c^\dagger_{\vec{k}\alpha} c_{\vec{k} \alpha'} + 2 \int \vec{\Phi}(\vec{x})  
\vec{S}(\vec{x}) d^3x
\end{equation}
where $\vec{S}(\vec{x})=\sum e^{i (\vec{k}-\vec{k}')\vec{x}}
c^\dagger_{\vec{k}\alpha} \, \vec{\sigma_{\alpha \alpha'}}/2 \, c_{\vec{k}'
  \alpha'}$ is the spin of the conduction electrons and
$\vec{\Phi}(\vec{x})$ is defined in eq.~(\ref{phi0}). Note that
(\ref{h}) is already strongly simplified: in real systems
$\epsilon_{\vec{k}}^{\alpha \alpha'}$ can depend 
on the magnetization and 
interband couplings may be relevant for a quantitative description. We believe, however, that the main {\em
  qualitative} effects are correctly captured by (\ref{h}).

In a first step, we neglect spin-orbit effects in the band-structure,
assuming that $\epsilon_{\vec{k}}^{\alpha \alpha'}=\epsilon^0_\vec{k}
\delta_{\alpha \alpha'}$.  While we will show below that this
approximation is {\em not} justified, it will serve as a starting
point for the full calculation. If the band-structure is spin-rotation
invariant, a residual symmetry leads to a very simple band-structure.
In the magnetically ordered state, the helical modulation breaks
spontaneously both the spin-rotation invariance and the translational
invariance along $\QQ$.  However, the product of a translation by a
lattice vector $\vec{x}$ and a rotation of spins around the $\QQ$ axis
by an angle $- \QQ \vec{x}$
\begin{equation}\label{sym}
\tilde{T}_\vec{x}= e^{i \vec{x} \vec{P}} e^{- i \QQ \vec{x} \int (\vec{S}(\vec{x}') \QQ)/q_0 \, d^3 x'}
\end{equation}
is still a symmetry of the Hamiltonian, where $\vec{P}$ is the
generator of translations.  As a consequence, instead of many
minibands (in a reduced Brillouin zone of width $q_0$) just two bands
form like in the ferromagnetic state. This can be seen by going to a
frame of reference rotating with the helix, i.e.  by applying the
unitary transformation $U$ with
\begin{equation} \label{U}
U=e^{i \int q_0 z S_z(\vec{x})
  \, d^3 x}, \quad U^+ c_{\vec{k},\sigma} U=c_{\vec{k}-\sigma \QQ/2,\sigma}
\end{equation}
to the Hamiltonian where we assumed for notational simplicity $\QQ \|
\hat{z}$.  In the rotated frame, $\Phi(x)=\Phi_0
\hat{x}$ is a constant vector pointing into the $\hat{x}$ direction
and the dispersion can be calculated from the $2\times 2$ matrix $
\left( \begin {array}{cc} \epsilon^0_{{\vec k} + \frac {\QQ}{2}} &
    \Phi_0 \\ \Phi_0 & \epsilon^0_{{\vec k} - \frac {\QQ}{2}} \end
  {array} \right)$ with eigenvalues
\begin{eqnarray} \label{2bands}
E_{\pm} ({\vec k}) &=& \frac{\epsilon^0_{{\vec k} + \frac {\QQ}{2}} + \epsilon^0_{{\vec k} - \frac {\QQ}{2}}}{2} 
\pm \sqrt{\frac{\left(\epsilon^0_{{\vec k} + \frac {\QQ}{2}} - 
\epsilon^0_{{\vec k} - \frac {\QQ}{2}}\right)^2}{4}+\Phi_0^2}
\approx  \epsilon^0_{\vec k} \pm \Phi_0 \pm \frac{(\vec{v}_{\vec{k}} \QQ)^2}{8 |\Phi_0|}
\end{eqnarray}
where we used that $q_0$ is small. One essentially obtains in this
rotated frame the Fermi surface of a ferromagnetic state, slightly
deformed in the $\QQ$ direction and -- as
emphasized above -- no minibands form.

\section{Full band structure} Due to spin-orbit coupling terms, real
systems are not invariant under the symmetry transformation
(\ref{sym}) as crystals are not rotationally invariant. It is
therefore essential to include also spin-orbit effects in the band
structure, $\epsilon_{\vec{k}}^{\alpha \alpha'}=\epsilon^0_{\vec{k}}
\delta_{\alpha \alpha'}+\sum g^i_{\vec{k}} \sigma^i_{\alpha \alpha'}$
where $\vec{g}_\vec{k}=(g^x_{\vec{k}},g^y_{\vec{k}},g^z_{\vec{k}})$
transforms like a vector under the
point group $T$ (i.e.  is a basis for the only 3 dimensional
irreducible representation of $T$), for example $\vec{g}_\vec{k}
\propto \vec{k}$ (compare to \cite{sigrist}, erratum).  As $\vec{g}_\vec{k}$ arises to first order in the
spin-orbit coupling, typical matrix elements are of order $\delta_B
\epsilon_F$, where $\epsilon_F$ is the Fermi energy and the
dimensionless constant $\delta_B$ has the same order of magnitude as
$\dDM$
\begin{equation}
\delta_B\sim \dDM=q_0/k_F \sim \delta
\end{equation}
and is therefore small, i.e. of the order of a few percent in MnSi.

We proceed by transforming first to the rotating frame of reference
using eq.~(\ref{U}) and then to the eigenstates $d_{\vec{k}, \pm}$
corresponding to the energies $E_{\pm}(\vec{k})$ in
eq.~(\ref{2bands}).  Using that close to the Fermi energies
$E_{\pm}(\vec{k})\approx 0$ and therefore $\epsilon^0_\vec{k}\approx
\mp \Phi_0$ we find $d_{\vec{k}, \pm}\approx \frac{1}{\sqrt{2}} (c_{
  \vec{k}\downarrow}\pm c_{ \vec{k}\uparrow})$ to lowest order.
Therefore the spin-orbit terms in the band-structure lead (for
$\QQ\|\hat{z}$) to a coupling of the form
\begin{equation}
H_{SO}^B\approx \frac{1}{2}\sum_\vec{k} 
\left(\!\begin{array}{l}
d^\dagger_{\vec{k}-\QQ/2,+}-d^\dagger_{\vec{k}-\QQ/2,-}\\
d^\dagger_{\vec{k}+\QQ/2,+}+d^\dagger_{\vec{k}+\QQ/2,-}
\end{array}\!
\right)^T \!
\left(\!\begin{array}{ll}
g^z_{\vec{k}} &g^x_{\vec{k}}- i g^y_{\vec{k}} \\
g^x_{\vec{k}}+i g^y_{\vec{k}} \ &  -g^z_{\vec{k}}
\end{array}\!
\right) 
\left(\!\begin{array}{l}
d_{\vec{k}-\QQ/2,+}-d_{\vec{k}-\QQ/2,-}\\
d_{\vec{k}+\QQ/2,+}+d_{\vec{k}+\QQ/2,-}
\end{array}\!
\right)
\end{equation}
and the total Hamiltonian is given by $H=H_s+H_{SO}^B$ with
$H_{s}=\sum_{i=\pm} E_i(\vec{k}) d^\dagger_{\vec{k}i} d_{\vec{k}i}$.
As $H_{SO}^B$ induces transitions from $\vec{k}$ to $\vec{k}+\QQ$ it
leads to the the formation of mini-bands in the new Brillouin zone of
width $q_0$. As $\QQ$ is small, transitions between the well separated
$+$ and $-$ band can be neglected and contributions due to
$g^z$ lead only to a minor deformation of the  bands.
For small $\delta_B$ and $\dDM$ the only remaining terms are 
\begin{eqnarray}\label{hso}
H_{SO}^{B,-}&=&\sum E_-(\vec{k}) d^\dagger_{\vec{k}-} d_{\vec{k}-}+\sum d^\dagger_{\vec{k}+\QQ/2,-} 
\frac{g^x_{\vec{k}}- i g^y_{\vec{k}}}{2} d_{\vec{k}-\QQ/2,-}+ h.c. \nonumber
\end{eqnarray}
and a similar contribution for the other band. The main corrections to
the band-structure will arise in regions of the Fermi surface where
$E_{\vec{k},-}-E_{\vec{k}+\QQ,-}$ is very small. We therefore expand
$E_-(\vec{k})$ around planes in momentum space where the Fermi
velocity $\partial_\vec{k} E_-(\vec{k})$ is perpendicular to $\QQ$ and
we introduce new momentum coordinates $(\kappa_\perp,\kappa_z,n)$,
where $n$ is an integer and $-q_0/2<\kappa_z\le q_0/2$ is a momentum in the appropriate
reduced Brillouin zone with $k_z=k_z^0+\kappa_z+n q_0$ and  $\partial_\vec{k} E_-(\kappa_\perp,k_z^0)
\QQ=0$ by construction. With these definitions we obtain
\begin{eqnarray}
E_-(\kappa_\perp,\kappa_z,n)&\approx&
E_-(\kappa_\perp)+(\kappa_z+n q_0)^2/(2 m_{\kappa_\perp})\\
g^x_{\vec{k}}- i g^y_{\vec{k}}&\approx& const.
\end{eqnarray}
where $m_{\kappa_\perp}$ is a measure of the curvature of the Fermi surface. For fixed $\kappa_\perp$ and $k_z$ the Hamiltonian (\ref{hso}) takes the form
\begin{eqnarray}\label{hso2}
H^{\kappa_\perp,\kappa_z}_{SO}\approx \epsilon_F \sum_{n} c_1 \delta_{B}   (d^\dagger_{n+1} d_{n}+d^\dagger_{n} d_{n+1})
 + c_2 \dDM^2  
(n-\frac{\kappa_z}{q_0})^2  d^\dagger_{n} d_{n}+c_3
\end{eqnarray}
where $c_1=|g^x_{k_\perp,k_z^0}- i g^y_{k_\perp,k_z^0}|/(2 \delta_B
\epsilon_F)$, $c_2=k_F^2/(2 m_{\kappa_\perp} \epsilon_F) $ and
$c_3=E_-(\kappa_\perp,k_z^0)/\epsilon_F$ are constants of order 1.
Note that it may happen that $|\vec{g}_\vec{k} \times \QQ|$, i.e.
$c_1$, vanishes at some points on the Fermi surface. This case will be
discussed separately below. For finite $c_1$, it is obvious from eq.~(\ref{hso2}) that
the contribution from $g^x_{\vec{k}}- i g^y_{\vec{k}}$ (the first
term), which is {\em linear} in $\delta_B$, will strongly modify the
band-structure of eq.~(\ref{2bands}) (the second term), which
contributes only to order $\dDM^2\sim \delta_B^2$. To analyze the
spectrum of (\ref{hso2}) it is useful to make a Fourier transform from
band-index space $n$ to the conjugate variable $\xi$ which leads to
\begin{equation}\label{hso3}
H^{\kappa_\perp}_{SO} \approx \epsilon_F \left[ -c_2 \dDM^2  
\partial_\xi^2  +  2 c_1 \delta_B \cos(\xi) +c_3 \right]
\end{equation}
where we switched to first-quantized language and absorbed the
dependence on $\kappa_z$ in the boundary condition $\Psi(\xi+2 \pi)=e^{i
  \xi \kappa_z/q_0 }\Psi(\xi)$ by a gauge transformation.
Eq.~(\ref{hso3}) is the familiar Hamiltonian of a particle in a
periodic potential and according to the boundary condition the
bandwidth in this model corresponds directly to the bandwidth in the
$\QQ$ direction in our original model. Similar models show up in many
different problems, see e.g. ref.~\cite{mauz} where
eigenfunctions of (\ref{hso2}) are discussed in detail.

\begin{figure}
\begin {tabular}{cccccc}
a) & \hspace{-3mm} \includegraphics[scale=0.3]{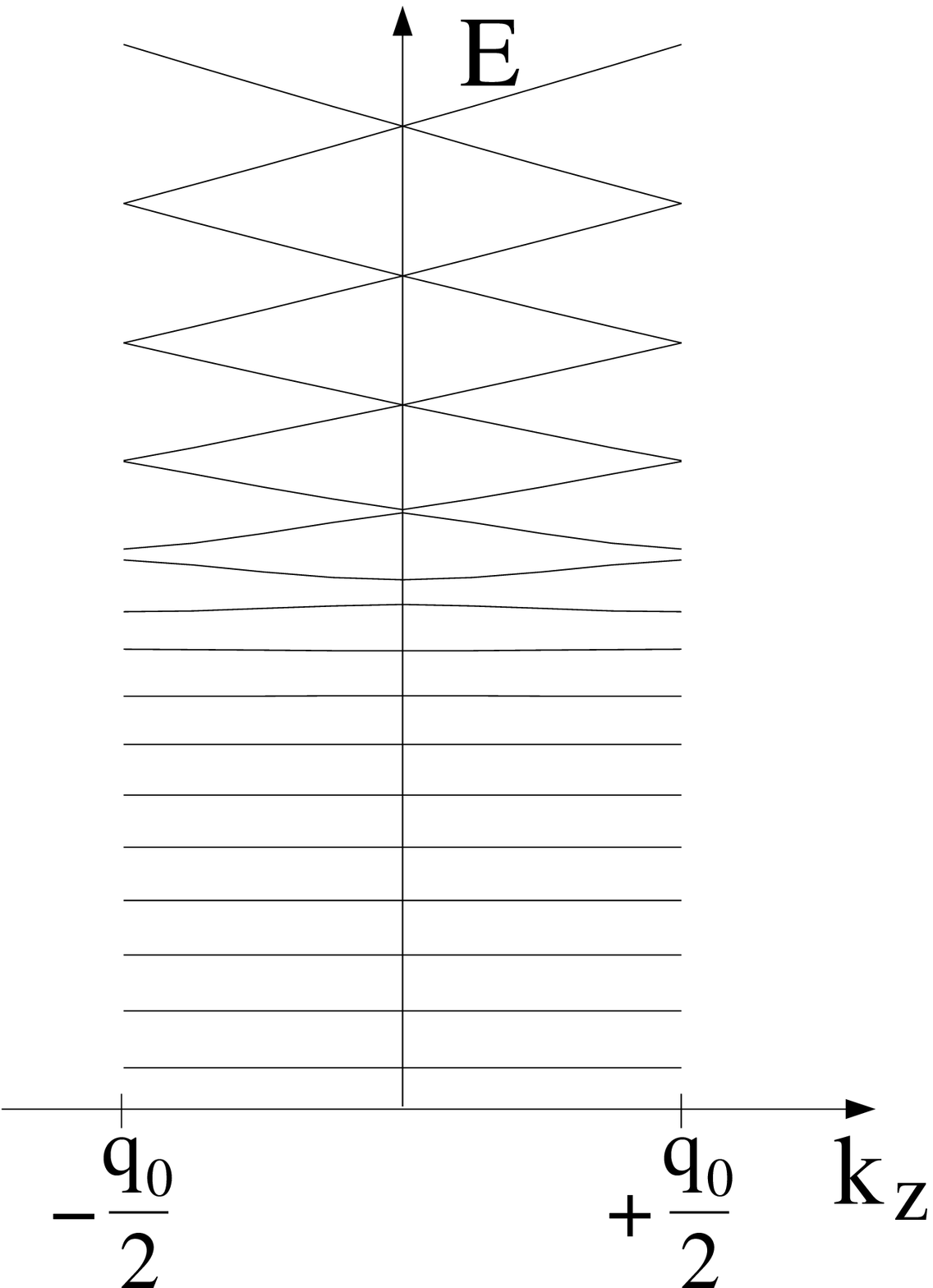} &
b) & \hspace{-3mm} \includegraphics[scale=0.3]{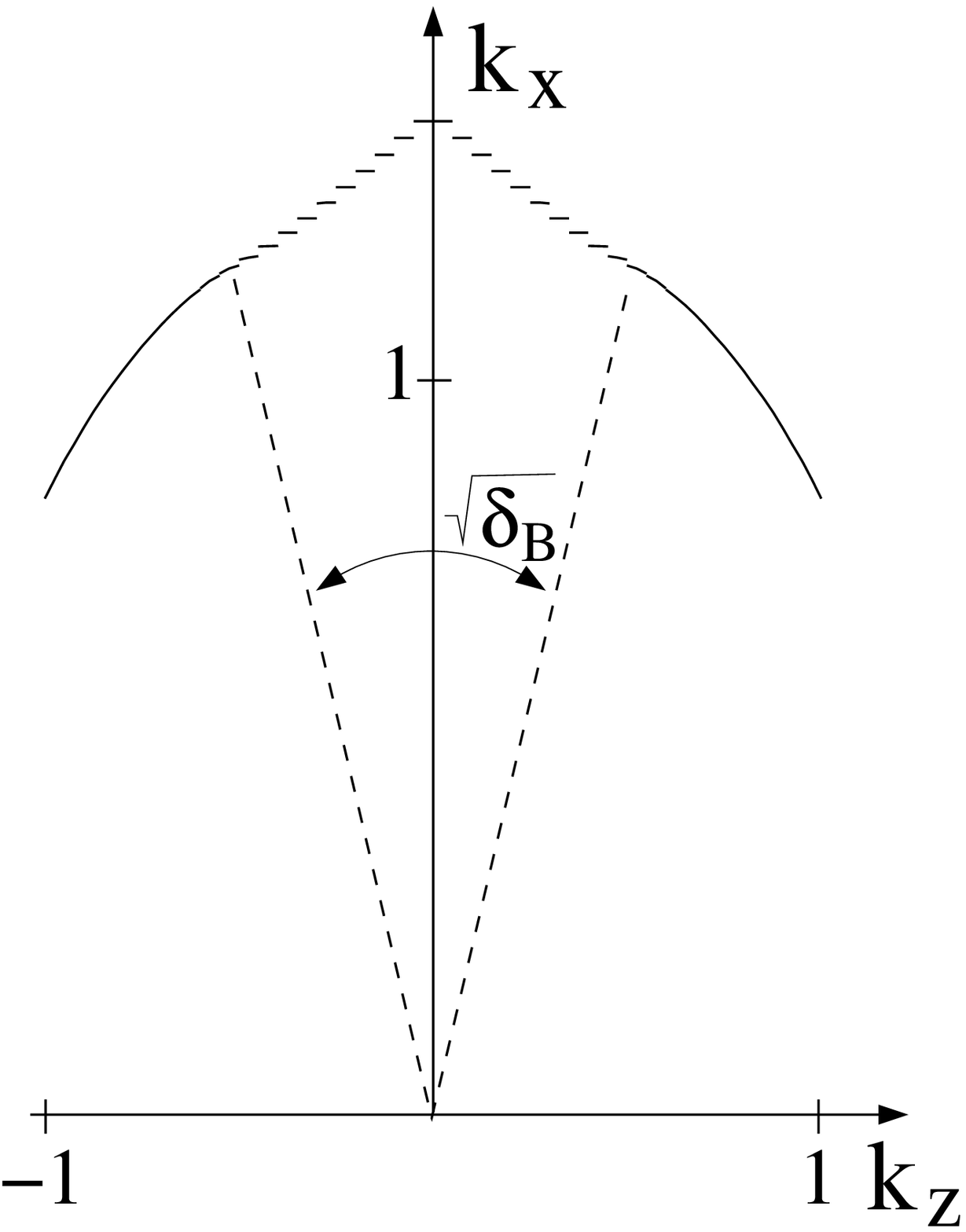}
\hspace{2mm} c) & \hspace{-3mm} \includegraphics[scale=0.15]{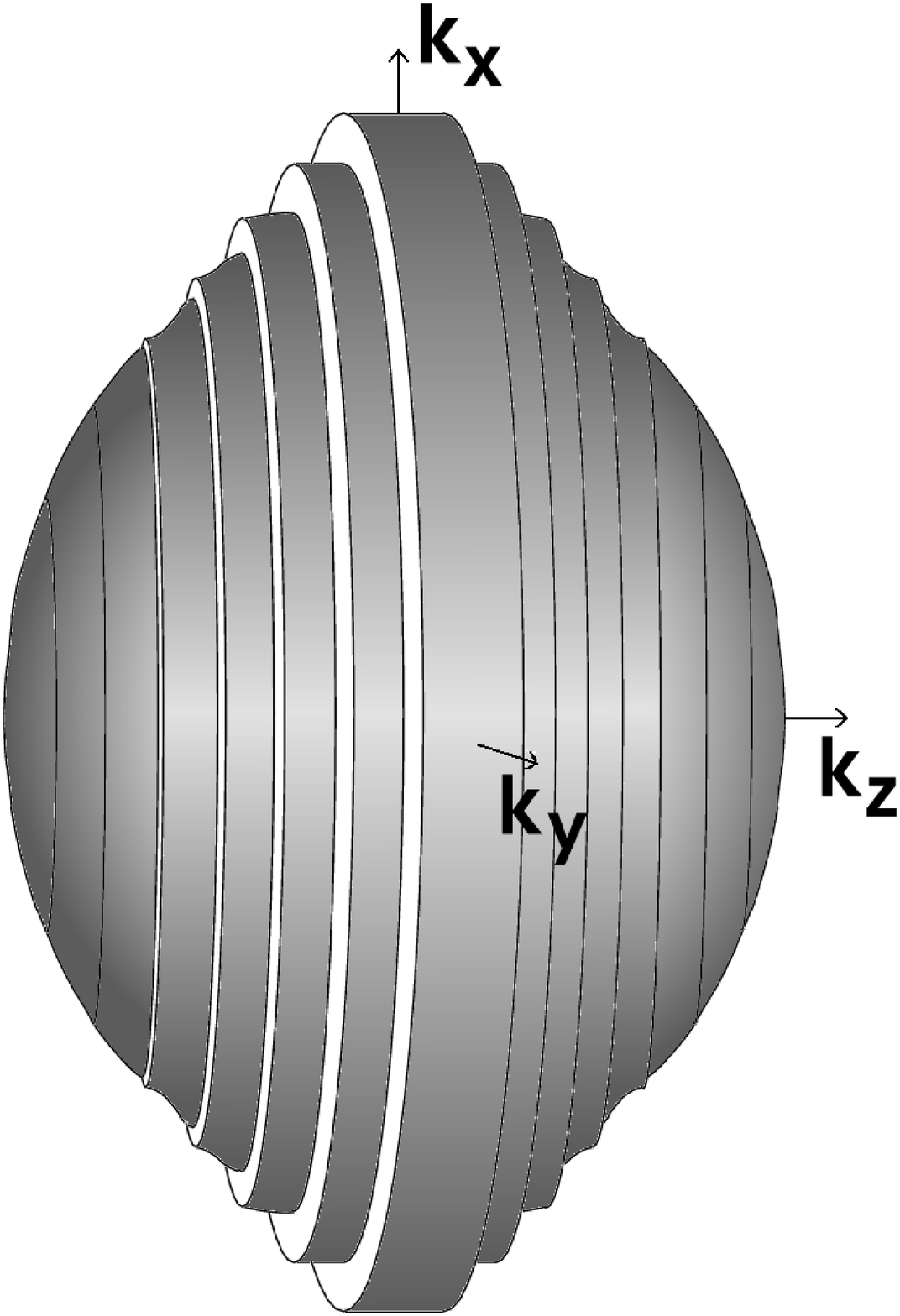}
\end {tabular}
\caption{ a) Band structure of the Hamiltonian (\ref{h}) with $\epsilon_\vec{k}^0=k^2/2-1/2$, $\Phi_0=0.3$, 
  $\QQ=(0,0,0.1)$, $\vec{g}_\vec{k}=0.1 \vec{k}$, $k_x =1.26$, $k_y=0$
  which corresponds to $\dDM=\delta_B=0.1$, $c_1 \approx 1.26$,
  $c_2\approx 1$. The lowest bands have an exponentially small band
  width, e.g. $2\cdot 10^{-10} \epsilon_F$ for the first band, consistent with eq.~(\ref{bandw}). b) Resulting Fermi surface for
  $k_y=0$ in an extended Brillouin zone. Only the $E_-$ bands are
  shown.  c) Sketch of the Fermi surface. In a belt of width
  $\sqrt{\delta_B}$ the mini-bands are completely flat in the
  direction parallel to the helix.
\label{fig1}}
\end{figure}

The spectrum of (\ref{hso3}) can be easily obtained in two limits.
For energies $E\gg E_0=\epsilon_F \delta_{B}$ much higher than the
cosine potential, the periodic potential has little influence. Up to
{\em exponentially} small and therefore irrelevant mini-band gaps, one
recovers the dispersion (\ref{2bands}). More important is the other
limit $E \ll E_0$, where the ''particle'' sits deep in the minima of
the cosine potential and is therefore well described by a harmonic
oscillator problem where the band-splitting is given by the harmonic
oscillator frequency
\begin{eqnarray}
\Delta E =2 \epsilon_F  \dDM\sqrt{\delta_B}  \sqrt{c_1 c_2}\propto \delta^{3/2}.
\end{eqnarray}
The finite band-width $W$ in $\QQ$ direction arises from the {\em
  exponentially} small tunneling from one minimum of the cosine to the
next which can be calculated with the help of a WKB approximation or an
instanton expansion \cite{coleman}
\begin{eqnarray}\label{bandw}
W \approx \frac{\Delta E}{\sqrt{\pi}} \exp\!\left[ - 8 \sqrt{\frac{c_1}{c_2}} \frac{\sqrt{\delta_B}}{\dDM} \right] \sim \delta^{3/2} \exp\!\left[ -c'\frac{1}{\sqrt{\delta}}\right]
\end{eqnarray}
This is the main result of our paper. The interplay of the weak
Dzyaloshinsky-Moriya interaction of strength $\dDM$ and weak
spin-orbit coupling in the band-structure parametrized by $\delta_B$
leads to the formation of exponentially flat bands.  In those bands,
the motion of the electrons parallel to $\QQ$ is practically not
possible. The non-analytic dependence of $W$ and $\Delta E$ on
$\delta_B$ and $\dDM$ shows that this is a non-perturbative effect
involving the interplay of resonant backscattering from the helix and
crystal field effects. Corrections to the exponent in
eq.~(\ref{bandw}) are of the order $(n+1/2)^2
\delta_{DM}/\sqrt{\delta_B}$, where $n$ is the band index. We have
confirmed our analytical results by comparing to a numerical
diagonalization of the Hamiltonian (\ref{h}) shown in fig.~\ref{fig1}.

What fraction of the Fermi surface is affected by those extremely flat
bands?  To answer this question, we count the number of bands
$N$ whose energy is smaller than $E_0$, $N\sim E_0/\Delta E \sim
\sqrt{\delta_B}/\dDM$. Therefore electrons on a belt on the Fermi
surface of width (see fig.~\ref{fig1})
\begin{eqnarray}
N q_0 \sim k_F \sqrt{\delta_B} 
\end{eqnarray}
will stop their motion parallel to $\QQ$. As expected the effect
vanishes in the limit $\delta_B \to 0$, we note however, that even for
very small $\delta_B$ of the order of a few percent, a sizable
fraction of the Fermi surface is affected by those ultra-flat bands.

It may happen that $|\vec{g}_{\vec{k}} \times \QQ|$ vanishes at some
singular points on the Fermi surface. If the lines where $\QQ$ is
parallel to the Fermi surface happen to be close by, our
results quoted above are not valid close to those points (but remain
valid elsewhere). In this case one can approximately linearize
$g^x_{\vec{k}}- i g^y_{\vec{k}}$ and replace the first
term in eq.~(\ref{hso2}) by $\epsilon_F \sum_n c_1' \dDM \delta_B
(d^\dagger_{n+1} d_{n}+d^\dagger_{n} d_{n+1}) (n-\kappa_z/q_0)$. As
this term is only of order $\delta^2$, only very few (of order
$\delta_B/\dDM\approx 1$) pronounced minibands will form which cover
close to that point only a tiny area of the Fermi surface of width
$k_F \delta_B$.

\section{Experimental Consequences} Unfortunately, the exponentially flat
minibands described above are not easily observed in a direct way --
with the exception of measurements of the anomalous skin effect. As
both the size of the Brillouin zone and the band width are so small, an
unrealistically clean sample with a mean free path of the order of
centimeters would be required to observe de-Haas-van-Alphen \cite{deHaas} or
Shubnikov-de-Haas oscillations. More indirectly, one could try to
observe changes in the (residual) resistivity due to the opening of
band gaps which stop the motion along $\QQ$. In the absence of the
helix, the electrons in the ''belt'' of width $\sqrt{\delta_B}$ shown
in fig.~\ref{fig1} have a velocity of order $v \sim \delta_B$
parallel to $\QQ$. Therefore their total contribution to transport
parallel to $\QQ$ is only of the order $v^2 \delta_B \sim
\delta_B^{3/2}$ and one expects changes of approximately this
magnitude when comparing transport parallel and perpendicular to the
helix. Again this effect is difficult to be observed.

A large effect can, however, be expected from any experimental
technique which probes only the electrons in the relevant region of
momentum space.  This is precisely the case in experiments measuring
the anomalous skin effect \cite{skin}. In sufficiently dirty metals with
conductivity $\sigma$ the penetration or skin depth of electromagnetic
waves of frequency $\omega$ is given by 
\begin{equation}
\Delta_0=\sqrt{2/(\sigma
  \omega \mu \mu_0)} \label{d0}
\end{equation}
where $\mu \mu_0$ is the magnetic permeability.  However, this
relation is only valid if the skin depth $\Delta$ is large compared to
the mean free path $l_0$.  In the other limit, for $\Delta < l_0$, the
so-called anomalous skin effect arises \cite{skin}: All electrons
which have a sufficiently large component $v_F^\perp$ of the velocity
perpendicular to the surface of the metal can escape the influence of
the electric field {\em before} scattering and only electrons moving
almost parallel to the surface of the crystal,
$v_F^\perp/v_F<\Delta/l_0$, contribute to the screening of the
electromagnetic wave. Note, that the minibands affect the motion of
electrons with a small component of the velocity parallel to the
direction of the helix. One can therefore expect a pronounced
modification of the skin depth when orienting the helix either
parallel or perpendicular to the surface of the metal. Fortunately, a
small magnetic field of about 0.12\,T is
sufficient\cite{kadowaki,plumer} to orient the helix parallel to the
field and by rotating the magnetic field one can determine the
direction of the helix.

If the helix is oriented {\em parallel} to the surface, $v_F^\perp$ is
not affected by the minibands. As only electrons with
$v_F^\perp/v_F<\Delta/l_0$ contribute to $\sigma$, the conductivity in
eq.~(\ref{d0}) is reduced by a factor of order $\Delta/l_0$ and
therefore the anomalous skin depth $\Delta_a^\|$ is estimated to be
\begin{equation}
\Delta_a^\| \sim \Delta_0 \sqrt{l_0/\Delta_a}\sim  (\Delta_0^2 l_0)^{1/3}.
\end{equation}
For an orientation of the helix {\em perpendicular} to the crystalline
surface all electrons in a belt of width $\sqrt{\delta_B}$ move
parallel to the surface, therefore the effective surface conductivity
will not drop below $\sigma \sqrt{\delta_B}$, which leads to a skin depth of
order
\begin{equation}
\Delta_a^\perp \sim \Delta_0/\delta_B^{1/4}
\end{equation}
for $\Delta_a^\|/l_0 \ll \sqrt{\delta_B}$ or $\Delta_0 \ll l_0
\delta_B^{3/4}$. By rotating the small external magnetic field one can
directly compare the two orientations experimentally,
\begin{equation}\label{ratio}
\Delta_a^\| / \Delta_a^\perp \sim \delta_B^{1/4} 
\left(\frac{l_0}{\Delta_0}\right)^{1/3}.
\end{equation}
While it may be difficult to observe large ratios, a sizable
qualitative effect can be expected for $\Delta_0/l_0 \lesssim
\delta_B^{3/4}$. MnSi crystals can be grown with exceptionally high
quality, ref.~\cite{lonzarich} reports a residual resistivity
$\rho_0\approx 0.17\,\mu\Omega$cm.  For frequencies of for example
20\,GHz one obtains $\Delta_0\approx 1500\,$\AA \ which is already
well below the estimated \cite{lonzarich} mean-free path,
$l_0>5000\,$\AA.  While larger frequencies or cleaner samples
($\Delta_0/l_0 \propto \rho_0^{3/2}/\sqrt{\omega}$) may be needed to
reach the regime $\Delta_0/l_0 <
\delta_B^{3/4}$ where eq.~(\ref{ratio}) is valid, we nevertheless
expect \cite{private} a measurable dependence on the field orientation
for the parameters used above. A more quantitative estimate would
require a detailed study of the band structure.

\section{Conclusions}
We have shown that two small relativistic effects of spin-orbit
coupling in a crystal without inversion symmetry, the
Dzyaloshinsky-Moriya interaction and the splitting of the Fermi
surface for electrons of opposite spin, conspire to produce
exponentially flat minibands in a magnetic metal. These minibands
induce pronounced variations in the anomalous skin depth for example
in clean MnSi crystals when the helical magnetic state is rotated by a
small magnetic field.  In this paper we have considered only simple
band-structure effects arising in a mean-field description of the
magnetic phase. Presently it is an open question whether and how this
physics will also affect the inelastic scattering of electrons
\cite{roschToBe} or the nature of the phase
transition\cite{schmalian}.  This is especially interesting in the
high-pressure phase of MnSi where on the one hand the anomalous
temperature dependence of the resistivity $\Delta \rho \sim T^{1.5}$
points towards the existence of a highly unconventional scattering
mechanism and where on the other hand helical order
seems to survive  on long (but not infinite) time and length scales
\cite{nature}.

\acknowledgments
We acknowledge useful discussions with D.~Broun, W.~Hardy, J.~Schmalian, N.~Shah, M.~Sigrist, E.~Tsitsishvili, M.~Vojta,
P.~W\"olfle and especially M.~Garst and C. Pfleiderer.  Part of this work
was supported by SFB 608 and the Emmy-Noether program of the DFG.

\end{document}